\def\revtex{1}

\ifx\revtex\undefined

\documentclass[entropy,article,submit,oneauthor,pdftex]{Definitions/mdpi}

\firstpage{1}
\pubvolume{1}
\issuenum{1}
\articlenumber{0}
\pubyear{2021}
\copyrightyear{2021}
\datereceived{}
\dateaccepted{}
\datepublished{}
\hreflink{https://doi.org/} 

\Title{Multistatic radar measurements of Unidentified Aerial Phenomena by cell and open access radio networks}

\TitleCitation{Open access multistatic radar measurements}

\Author{Karl Svozil $^{1}$\orcidA{}}

\AuthorNames{Karl Svozil}

\AuthorCitation{Svozil, K.}

\address[1]{%
$^{1}$ \quad Institute for Theoretical Physics, TU Wien, Wiedner Hauptstrasse 8-10/136, 1040 Vienna,  Austria; svozil@tuwien.ac.at; \url{http://tph.tuwien.ac.at/~svozil}}

\corres{Correspondence: svozil@tuwien.ac.at}

\abstract{Measurements of a wide variety of aerial phenomena such as drones, birds, planes, and other aerial objects can in principle be obtained from open access radio detection and ranging equipment. Passive strategies involve existing cell towers as transmitters and cell phones as receivers. Another active possibility is the creation of a multistatic, possibly phased array, network of transmission and receiving stations by open-source realizations such as GNU radio. We also suggest provision of uncorrelated targets or anomalies from existing military sources.}

\keyword{Passive radar detection active radar detection, Unidentified Aerial Phenomena, Unidentified Aerial Vehicle, Unidentified Flying Object, Multistatic radar}

\keyword{Passive radar detection active radar detection, Unidentified Aerial Phenomena, Unidentified Aerial Vehicle, Unidentified Flying Object, Multistatic radar}

\DeclareFontFamily{U}{bbold}{}
\DeclareFontShape{U}{bbold}{m}{n}
 {
  <-5.5> s*[1.069] bbold5
  <5.5-6.5> s*[1.069] bbold6
  <6.5-7.5> s*[1.069] bbold7
  <7.5-8.5> s*[1.069] bbold8
  <8.5-9.5> s*[1.069] bbold9
  <9.5-11> s*[1.069] bbold10
  <11-15> s*[1.069] bbold12
  <15-> s*[1.069] bbold17
 }{}

\begin{document}

\else
\documentclass[%
      reprint,
   twocolumn,
 amsmath,amssymb,
 aps,
 pra,
  longbibliography,
 ]{revtex4-2}

\usepackage[dvipsnames]{xcolor}

\usepackage{mathptmx}

\usepackage{amssymb,amsthm,amsmath,bm}

\usepackage{tikz}
\usetikzlibrary{calc,decorations.pathreplacing,decorations.markings,positioning,shapes,snakes}

\usepackage[breaklinks=true,colorlinks=true,anchorcolor=blue,citecolor=blue,filecolor=blue,menucolor=blue,pagecolor=blue,urlcolor=blue,linkcolor=blue]{hyperref}
\usepackage{graphicx}
\usepackage{url}

\ifxetex
%
%
\usepackage{fontspec}
\usepackage{fontspec}
\setmainfont{Garamond}
\setsansfont{Garamond}
\fi

\usepackage{mathbbol} 

\begin{document}

\title{Multistatic radar measurements of Unidentified Aerial Phenomena by cell and open access radio networks}

\author{Karl Svozil}
\email{svozil@tuwien.ac.at}
\homepage{http://tph.tuwien.ac.at/~svozil}

\affiliation{Institute for Theoretical Physics,
TU Wien,
Wiedner Hauptstrasse 8-10/136,
1040 Vienna,  Austria}

\date{\today}

\begin{abstract}
Measurements of a wide variety of aerial phenomena such as drones, birds, planes, and other aerial objects can in principle be obtained from open access radio detection and ranging equipment. Passive strategies involve existing cell towers as transmitters and cell phones as receivers. Another active possibility is the creation of a multistatic, possibly phased array, network of transmission and receiving stations by open-source realizations such as GNU radio. We also suggest provision of uncorrelated targets or anomalies from existing military sources.
\end{abstract}

\keywords{Passive radar detection active radar detection, Unidentified Aerial Phenomena, Unidentified Aerial Vehicle, Unidentified Flying Object, Multistatic radar}

\maketitle

\fi

\section{Open source radar}

What and who resides and moves in the earth's atmosphere is of considerable interest to individuals, the public at large, as well as to state agencies.
It might not be too speculative to assume that the latter group, in particular signal intelligence and military,
is capable of monitoring and analyzing such activities with a wide variety of formidable sensors and automated data integration tools.
But for multiple reasons, even civil agencies---which
are often incorporated monopolies difficult to oblige with freedom-of-information requests---are
reluctant to share raw or processed sensor data, say, in the form of standardized interfaces.

To satisfy public open access demand for data on atmospheric occupancy and maneuver, we, therefore, suggest creating a participative
network of standardized, open access, multipartite radio detection, and ranging equipment.
Similar but maybe more ``passive'' projects have been realized for real-time lightning maps~\cite{blitzortung}
as well as unfiltered flight data air traffic monitoring~\cite{adsbexchange}.

We suggest ``free ride'' existing radio emitters, as well as creating a scalable low-cost adaptive network of emitters.
We also suggest utilizing available ``low-end'' receivers such as mobile phones.
The power of such a system is in its scalability and its resilience for all kinds of adverse interventions.
Its disadvantages may be seen in its low specificity for particular tasks---it may be perceived as the antipode of
highly effective and specialized phased array radar such as  AN/SPY-1---its high demand for computing power for signal analysis and interpretation,
a ``m\'elange of signal noise''~\cite{Chant},
as well as on its dependence on communication among parties.

\section{Passive cell phone-based radar}

In what follows we shall discuss passive configurations that presuppose transmitters as given and use receivers only.
The principles behind the idea to use cell towers as emitters and cell phones as receivers for radio detection and ranging are not entirely new.
Indeed, historically, passive, bi-static radar~\cite{Jackson1986,Willis-bs,Griffiths,Willis-InSkolnik,Willis-Griffiths-bs}
was the first type of radar used already in the mid-30th of the 20th century.
Passive radar, or, by another wording, passive coherent location (PCL) radar
utilize signals of opportunity, such as mobile, radio, or TV stations~\cite{Mathis-PCLR}.

Methods for real-time signal selection for passive coherent location systems can be found in a dissertation~\cite{Johnson-PCLS}.
A brief theoretical discussion and proof of concept realization of a bi-static system with one cell tower and one mobile phone
has been reported from a Swedish project~\cite{Cardin}.
A GSM-based PCL system for maritime surveillance has been implemented by
a group of the Fraunhofer Institute for Communications, Information Processing and Ergonomics~\cite{Zemmari2014,Chant}.

We shall now turn our attention to multi-static radar.
Contrary to bi-static radar we shall thereby assume that
\begin{itemize}
\item[(i)]
there are multiple, equally distributed transmitters over an infinite, flat plane; with an average distance $d_T$
\item[(ii)]
the horizontal height $d_O$ of the object $O$ is defined as the (minimal) distance
of the object from the ground plane is much larger than $d_T$;
that is, $d_T \ll d_O$;
\item[(iii)]
the object reflects radio signals equally in all directions.
\end{itemize}

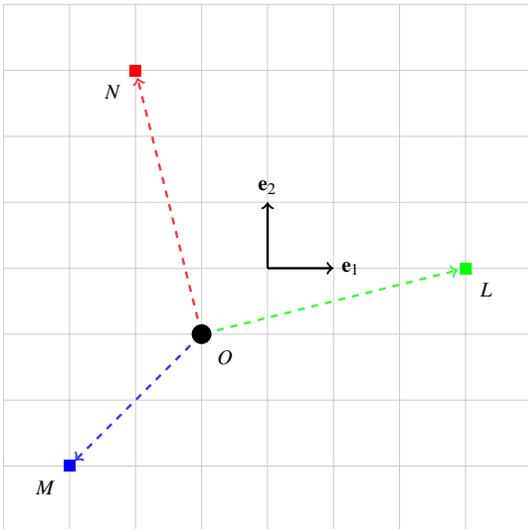
\begin{figure}
\begin{center}%
\resizebox{0.4\textwidth}{!}{
\begin{tikzpicture}  [scale=1]

\tikzstyle{every path}=[line width=1pt]

\newdimen\ms
\ms=0.1cm
\tikzstyle{s1}=[color=red,rectangle,inner sep=2.5]
\tikzstyle{c3}=[circle,inner sep={\ms/8},minimum size=4*\ms]
\tikzstyle{c2}=[circle,inner sep={\ms/8},minimum size=3*\ms]
\tikzstyle{c1}=[circle,inner sep={\ms/8},minimum size=2*\ms]
\tikzstyle{cs1}=[circle,inner sep={\ms/8},minimum size=1*\ms]


\coordinate (zero) at (0,0);
\coordinate (M) at (-3,-3);
\coordinate (N) at (-2,3);
\coordinate (L) at (3,0);
\coordinate (O) at (-1,-1);


 \draw[thin,gray!40] (-4,-4) grid (4,4);


 \draw[<->,shorten >=1.2mm, dashed, line width=1pt,blue!80] (O)--(M);
 \draw[<->,shorten >=1.2mm, dashed, line width=1pt,red!80] (O)--(N);
 \draw[<->,shorten >=1.2mm, dashed, line width=1pt,green!80] (O)--(L);


\draw (M) coordinate[s1,fill=blue,label=below left:{$M$}];
\draw (N) coordinate[s1,fill=red,label=below left:{$N$}];
\draw (L) coordinate[s1,fill=green,label=below right:{$L$}];
\draw (O) coordinate[c2,fill=black,label=below right:{$O$}];


  \draw[->] (0,0)--(1,0) node[right]{$\mathbf{e}_1$};
  \draw[->] (0,0)--(0,1) node[above]{$\mathbf{e}_2$};

\end{tikzpicture}
}
\end{center}
\caption{\label{2022-radar-conf}
Depiction of a normal projection onto a plane.
Three detectors $M,N,L$ are in that plane, and an object $O$ is above it at (minimal) distance $d_O$.
Transmitters are not drawn.}
\end{figure}

Suppose three receivers $M, N, L$ are located in the plane, with coordinates
$
\begin{pmatrix}
m_1,m_2,m_3=0
\end{pmatrix}^\intercal
$,
$
\begin{pmatrix}
n_1,n_2,n_3=0
\end{pmatrix}^\intercal
$,
$
\begin{pmatrix}
l_1,l_2,l_3=0
\end{pmatrix}^\intercal
$
relative to some orthonormal coordinate (basis) system $\{ {\bf e_1},{\bf e_2},{\bf e_3}\}$
whose first two elements ${\bf e_1}$ and ${\bf e_2}$ span the plane.
The symbol ``$\intercal$'' indicates transposition.
Any point in that plane has zero third coordinate.
Figure~\ref{2022-radar-conf} depicts such a configuration, with an elevated object $O$
with coordinates
$
\begin{pmatrix}
o_1,o_2,o_3=d_O
\end{pmatrix}^\intercal
$.

As a consequence of (i)--(iii), we may treat the object as a point charge radiating as a function of distance.
Thereby, the irradiation from multiple (equidistant) interfering radio transmitters serves as a sort of omnipresent ``electrosmog''
from the ground radiated upwards which, in first approximation, is equidistributed across the sky; just as an infinite plane charge creates a constant electric field.
Unlike the usual radar energy that expands during both the signal transmission and the reflected return,
resulting in inverse fourth power law, we shall assume an inverse-square
law---equivalent to the geometric dilution corresponding to point-source radiation into three-dimensional space---of
the energy or intensity $I_i$ received from the object at $i\in M,N,L$.
Wo obtain three equalities
\begin{equation}
\begin{split}
(m_1-o_1)^2 +
(m_2-o_2)^2 +
(o_3)^2
 = (I_M)^{-1},
\\
(n_1-o_1)^2 +
(n_2-o_2)^2 +
(o_3)^2
 = (I_N)^{-1},
\\
(l_1-o_1)^2 +
(l_2-o_2)^2 +
(o_3)^2
 = (I_L)^{-1}
\end{split}
\end{equation}
for the three unknowns  $o_1$, $o_2$, and $o_3=d_O$.
The case of less than three detectors is undetermined; four or more detectors allow cross-checks.

Here are some weaknesses of this method:
\begin{itemize}
\item[(i)]
the cell towers are often at elevated positions and emit mainly towards the ground;
so the upwards signal strength might be rather weak;
\item[(ii)]
more than one object reflect the radion causing considerable interference that might necessitate more stations
and more computing power to analyze the configuration;
\item[(iii)]
any variation from symmetric reflection by, say, the geometric shape of the object, may distort the signal.
\end{itemize}
Nevertheless, it might be worthwhile to demonstrate the capability of this method with some further proof of principle installations.

\section{Passive and active open access radio network}

Another option to get open access signal information might be by creating a network of
radio antennas and receivers.
Again, a passive installation uses only receiving stations. In what follows we shall use active radar schemes.

One candidate for such a realization is GNU radio, a free and open-source software development toolkit
that, according to its description~\cite{GNURadio}, provides signal processing blocks to implement software radios
that can be used with external radio-frequency hardware.

Multiple software radios can be bundled together, synchronized, and switched
according to a wide variety of purposes~\cite{Wyglinski2010}.
For instance, software radios could (cost) effectively implement phased arrays~\cite{VisserPA,HansenRC,Mailloux},
that is,
software-controlled arrays of antennas that create radio waves that can thereby be
steered to point in a variety of directions without moving the antennas~\cite{Tosovsky2009,Fennelly}.
The antennas do not mechanically move, but they are synchronized in such a way that, through interference,
their collective radiation forms wavefronts ``as if'' emitted from a single radio source.
The radiation direction depends on the synchronization, but can in principle extend over the entire sky.

In principle, such arrays can be extended ``below the horizon'' to cover an entire globe.
This can, for instance, be achieved by ``stitching'' (slightly) overlapping phased arrays; very similar to panorama photography.

\begin{figure}
\begin{center}%
\resizebox{0.48\textwidth}{!}{
\begin{tabular}{c}
\begin{tikzpicture}  [scale=1]

\tikzstyle{every path}=[line width=1pt]

\newdimen\ms
\ms=0.1cm
\tikzstyle{s1}=[color=red,rectangle,inner sep=2.5]
\tikzstyle{c3}=[circle,inner sep={\ms/8},minimum size=4*\ms]
\tikzstyle{c2}=[circle,inner sep={\ms/8},minimum size=3*\ms]
\tikzstyle{c1}=[circle,inner sep={\ms/8},minimum size=2*\ms]
\tikzstyle{cs1}=[circle,inner sep={\ms/8},minimum size=1*\ms]


\coordinate (zero) at (0,0);
\coordinate (O) at (0,3);

\def\w{0.6cm};
\def\d{0.4cm};
\def\g{2cm};

\coordinate (A11) at ({-\d/2-\w-\g},0);
\coordinate (A12) at ( {\d/2-\w-\g},0);
\coordinate (A21) at ({-\d/2-\g},0);
\coordinate (A22) at ( {\d/2-\g},0);
\coordinate (A31) at ({-\d/2+\w-\g},0);
\coordinate (A32) at ( {\d/2+\w-\g},0);

\coordinate (B11) at ({-\d/2-\w},0);
\coordinate (B12) at ( {\d/2-\w},0);
\coordinate (B21) at ({-\d/2},0);
\coordinate (B22) at ( {\d/2},0);
\coordinate (B31) at ({-\d/2+\w},0);
\coordinate (B32) at ( {\d/2+\w},0);

\coordinate (C11) at ({-\d/2-\w+\g},0);
\coordinate (C12) at ( {\d/2-\w+\g},0);
\coordinate (C21) at ({-\d/2+\g},0);
\coordinate (C22) at ( {\d/2+\g},0);
\coordinate (C31) at ({-\d/2+\w+\g},0);
\coordinate (C32) at ( {\d/2+\w+\g},0);


\draw[->] ({-1.7*\g},0)--({1.7*\g},0) node[right]{$\mathbf{e}_1$};

\draw[line width=3pt,red!80] (A11) -- node[midway,below]{$A_1$} (A12);
\draw[line width=3pt,red!80] (A21) -- node[midway,below]{$A_2$} (A22);
\draw[line width=3pt,red!80] (A31) -- node[midway,below]{$A_3$} (A32);

\draw[line width=3pt,blue!80] (B11) -- node[midway,below]{$B_1$} (B12);
\draw[line width=3pt,blue!80] (B21) -- node[midway,below]{$B_2$} (B22);
\draw[line width=3pt,blue!80] (B31) -- node[midway,below]{$B_3$} (B32);

\draw[line width=3pt,green!80] (C11) -- node[midway,below]{$C_1$} (C12);
\draw[line width=3pt,green!80] (C21) -- node[midway,below]{$C_2$} (C22);
\draw[line width=3pt,green!80] (C31) -- node[midway,below]{$C_3$} (C32);

\draw (O) coordinate[c2,fill=black,label=above:{$O$}];

\coordinate (AA1) at ({-\w-\g},1.5);
\coordinate (AA2) at ({-\g},1.5);
\coordinate (AA3) at ({\w-\g},1.5);

\coordinate (BB1) at ({-\w},1.5);
\coordinate (BB2) at (0,1.5);
\coordinate (BB3) at ({+\w},1.5);

\coordinate (CC1) at ({-\w+\g},1.5);
\coordinate (CC2) at ({+\g},1.5);
\coordinate (CC3) at ({\w+\g},1.5);

\draw[thick,red] ([shift={(AA1)}]20:{\d}) arc[radius={\d}, start angle=20, end angle= 160];
\draw[thick,red] ([shift={(AA2)}]20:{\d}) arc[radius={\d}, start angle=20, end angle= 160];
\draw[thick,red] ([shift={(AA3)}]20:{\d}) arc[radius={\d}, start angle=20, end angle= 160];

\draw[thick,blue] ([shift={(BB1)}]20:{\d}) arc[radius={\d}, start angle=20, end angle= 160];
\draw[thick,blue] ([shift={(BB2)}]20:{\d}) arc[radius={\d}, start angle=20, end angle= 160];
\draw[thick,blue] ([shift={(BB3)}]20:{\d}) arc[radius={\d}, start angle=20, end angle= 160];

\draw[thick,green] ([shift={(CC1)}]20:{\d}) arc[radius={\d}, start angle=20, end angle= 160];
\draw[thick,green] ([shift={(CC2)}]20:{\d}) arc[radius={\d}, start angle=20, end angle= 160];
\draw[thick,green] ([shift={(CC3)}]20:{\d}) arc[radius={\d}, start angle=20, end angle= 160];

\end{tikzpicture}
\\
(i)
\\
\\
\begin{tikzpicture}  [scale=1]

\tikzstyle{every path}=[line width=1pt]

\newdimen\ms
\ms=0.1cm
\tikzstyle{s1}=[color=red,rectangle,inner sep=2.5]
\tikzstyle{c3}=[circle,inner sep={\ms/8},minimum size=4*\ms]
\tikzstyle{c2}=[circle,inner sep={\ms/8},minimum size=3*\ms]
\tikzstyle{c1}=[circle,inner sep={\ms/8},minimum size=2*\ms]
\tikzstyle{cs1}=[circle,inner sep={\ms/8},minimum size=1*\ms]


\coordinate (zero) at (0,0);
\coordinate (O) at (0,3);

\def\w{0.6cm};
\def\d{0.4cm};
\def\g{2cm};

\coordinate (A11) at ({-\d/2-\w-\g},0);
\coordinate (A12) at ( {\d/2-\w-\g},0);
\coordinate (A21) at ({-\d/2-\g},0);
\coordinate (A22) at ( {\d/2-\g},0);
\coordinate (A31) at ({-\d/2+\w-\g},0);
\coordinate (A32) at ( {\d/2+\w-\g},0);

\coordinate (B11) at ({-\d/2-\w},0);
\coordinate (B12) at ( {\d/2-\w},0);
\coordinate (B21) at ({-\d/2},0);
\coordinate (B22) at ( {\d/2},0);
\coordinate (B31) at ({-\d/2+\w},0);
\coordinate (B32) at ( {\d/2+\w},0);

\coordinate (C11) at ({-\d/2-\w+\g},0);
\coordinate (C12) at ( {\d/2-\w+\g},0);
\coordinate (C21) at ({-\d/2+\g},0);
\coordinate (C22) at ( {\d/2+\g},0);
\coordinate (C31) at ({-\d/2+\w+\g},0);
\coordinate (C32) at ( {\d/2+\w+\g},0);


\draw[->] ({-1.7*\g},0)--({1.7*\g},0) node[right]{$\mathbf{e}_1$};

\draw[line width=3pt,red!80] (A11) -- node[midway,below]{$A_1$} (A12);
\draw[line width=3pt,red!80] (A21) -- node[midway,below]{$A_2$} (A22);
\draw[line width=3pt,red!80] (A31) -- node[midway,below]{$A_3$} (A32);

\draw[line width=3pt,blue!80] (B11) -- node[midway,below]{$B_1$} (B12);
\draw[line width=3pt,blue!80] (B21) -- node[midway,below]{$B_2$} (B22);
\draw[line width=3pt,blue!80] (B31) -- node[midway,below]{$B_3$} (B32);

\draw[line width=3pt,green!80] (C11) -- node[midway,below]{$C_1$} (C12);
\draw[line width=3pt,green!80] (C21) -- node[midway,below]{$C_2$} (C22);
\draw[line width=3pt,green!80] (C31) -- node[midway,below]{$C_3$} (C32);

\draw (O) coordinate[c2,fill=black,label=above:{$O$}];

\coordinate (AA1) at ({-\w-\g},1.3);
\coordinate (AA2) at ({-\g},1);
\coordinate (AA3) at ({\w-\g},0.7);

\coordinate (BB1) at ({-\w},0.6);
\coordinate (BB2) at (0,0.5);
\coordinate (BB3) at ({+\w},0.6);

\coordinate (CC1) at ({-\w+\g},0.7);
\coordinate (CC2) at ({+\g},1);
\coordinate (CC3) at ({\w+\g},1.3);

\draw[thick,red] ([shift={(AA1)}]20:{\d}) arc[radius={\d}, start angle=20, end angle= 160];
\draw[thick,red] ([shift={(AA2)}]20:{\d}) arc[radius={\d}, start angle=20, end angle= 160];
\draw[thick,red] ([shift={(AA3)}]20:{\d}) arc[radius={\d}, start angle=20, end angle= 160];

\draw[thick,blue] ([shift={(BB1)}]20:{\d}) arc[radius={\d}, start angle=20, end angle= 160];
\draw[thick,blue] ([shift={(BB2)}]20:{\d}) arc[radius={\d}, start angle=20, end angle= 160];
\draw[thick,blue] ([shift={(BB3)}]20:{\d}) arc[radius={\d}, start angle=20, end angle= 160];

\draw[thick,green] ([shift={(CC1)}]20:{\d}) arc[radius={\d}, start angle=20, end angle= 160];
\draw[thick,green] ([shift={(CC2)}]20:{\d}) arc[radius={\d}, start angle=20, end angle= 160];
\draw[thick,green] ([shift={(CC3)}]20:{\d}) arc[radius={\d}, start angle=20, end angle= 160];

\end{tikzpicture}
\\
(ii)
\end{tabular}
}
\end{center}
\caption{\label{2022-radar-al}
A sectional view of a schema for the adaptive location of an object by an extended array consisting of three subarrays in two dimensions.
Each of the three arrays consists of three antennae $A_1,A_2,A_3$, $B_1,B_2,B_3$  and  $C_1,C_2,C_3$
that (i) in the first phase emit wave fronts that are all upwards directed, and upon detection of object $O$
(ii) ``zero in'' on $O$ to obtain maximal intensity of the reflected signal.}
\end{figure}
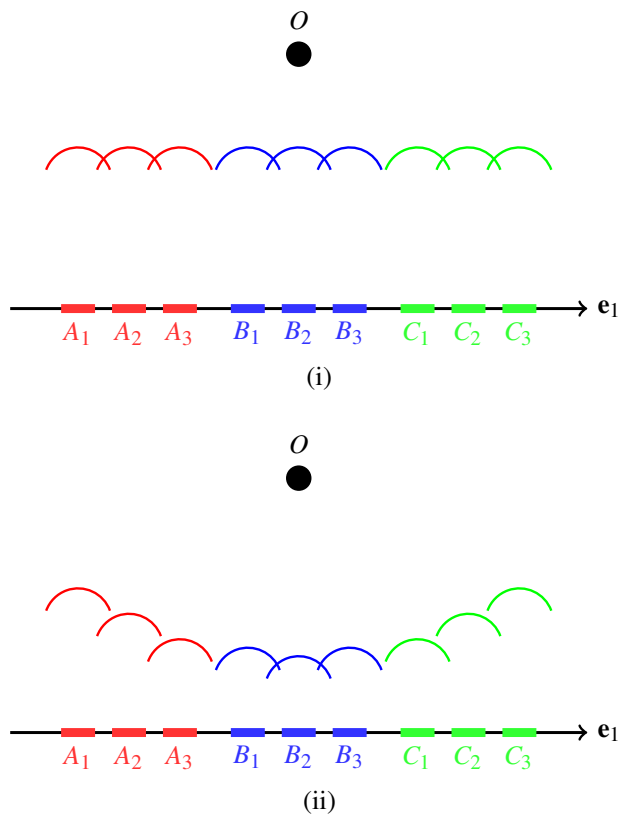

One way of utilizing the great plasticity and flexibility of these networks is the adaptive concentration on potential objects
my a very large array, covering large parts of the plane, or by two or more arrays ``chasing'' the object.
If, for instance, analysis shows a potential candidate for observation, the array may be adaptively switched in ways to improve detectability
and resolution.
Figure~\ref{2022-radar-al} depicts the schematics of an array consisting of two subarrays adaptively locating a flying object.

The radio signals need to be limited to the legally allowed radio frequencies~\cite{Roberson2010,USFAC} and signal amplitudes.
Such regulatory constraints vary.

\section{Open access signal autonomy and its limits}

\subsection{Open access radar initiatives}

The aforementioned proposals are directed towards greater autonomy of civil science and the public at large
through the open access
to certain types of information about the habitation of, and movement and signals in, earth's atmosphere.
There are more than 50 previous, current, and planned projects which scan ``the skies'' to some extent~\cite{Isaac-UAP-Trackers-list}; among them,
in lexicographic order,
the Galileo Project~\cite{GalileoProject},
HawkEye360~\cite{HawkEye360},
{Sky360}~\cite{Sky360} (the successor of SkyHub~\cite{SkyHub}),
SkyCAM-5~\cite{SkyCAM-5},
UAP Tracker~\cite{UAPTracker}, and
UAPx~\cite{UAPx.space}.

{\it Weak Signal Communication, by K1JT} (WSJT)~\cite{WSJT},
and its subcomponent {\it Weak Signal Propagation Reporter} (WSPR) (pronounced ``whisper''
provides an open, publicly available record of signals from
``network of transmitter beacons and receivers'')~\cite{wsprnet}.
For smaller distances---up to maximally 300-400 km---and in regions of ``good'' coverage~\cite[t=413s=6:53m]{HamRadioDX}.

\subsection{Open access to uncorrelated target categories at military installations: UTC/UTR/UER/UAEs}

\subsubsection{Category formation}
Tracking information could, in principle, be obtained with a much higher signal-to-noise ratio
from other than open sources, such as military and signals intelligence (SIGINT).
Historically, Vall\'ee recalled Sagan~\cite[t=4155s=1:09:15h]{Vallee-TJRE2020} asking the North American Aerospace Defense Command (NORAD)
if they registered any unidentified flying objects (UFOs). The answer was that, at that time (post-Blue Book), NORAD did not record any UFOs.
Nevertheless, at the same time Sagan was told that NORAD allegedly registered 10.000 so-called ``uncorrelated targets'' (UTC) per month.

A letter from NORAD clarifies the terminology by stating that~\cite{DeanPaul-NORAD}
``NORAD uses the term ``Unknown Track Report'' (UTR) for events within our
atmosphere and the United States Space Command (USSPACECOM) uses the term
``Uncorrelated Events'' for objects in space. UTR's are considered sensitive and not
releasable to the public. Uncorrelated Event Reports (UER's) are classified SECRET
until downgraded by proper authority. The term ``UFO'' has not been used by this
headquarters since the ``Blue Book'' was permanently closer in 1974.''

This brings up the fascinating issue of ``category formation'' or ``object construction'': What should be considered a noteworthy category,
and what kind of signals should be categorized as an ``anomaly'' or uncorrelated?
For the sake of coping with the immense amount of information gathered by physical sensors
these raw data must be subjected to a variety of categorization filters.
Primary categories of interest include (swarms of) intercontinental ballistic missiles or adversary craft of any kind, posing threats to defense.
Anomalies---that is,  objects with ``strange behavior'' or flight patterns that are impossible
with current physical means; in particular, abrupt accelerations or ``zig-zak''-style changes~\cite{Knuth-e21100939}---fall into the category of UTRs (or UERs).

Civil aviation authorities or some European military consider only of correlated, identifyable tracks.
They are not concerned with uncorrelated flight paths that
appear ``discontinuous''~\cite{Mueller-Radar-blind} or accelerate abruptly~\cite[Chapter~8]{VonLudwigerNIDS}.
Effectively anomalies such NORADs UTRs or USSPACECOMs UERs are implicitly excluded from consideration;
their signals are dropped, and the respective integrated systems turn intentionally blind on them.

And yet, it would be exactly those dropped uncorrelated targets that might require further investigation.
Given their potential importance for oversight of homeland security it may not be unreasonable
to speculate that they are classified and not dropped and deleted, and are continuously recorded and reported to unknown secret facilities.

\subsubsection{Open access to uncorrelated targets}

One may argue that, as these uncorrelated target categories allegedly do not present any threat to homeland security~\cite{Condon-report} they might,
after suitable ``degradation'' for unclassified civil use, be made openly available~\cite{Dean-Willis-podcast-2020,Mellon2022Feb}.

At first sight the public access of such uncorrelated target categories from military and SIGINT data appears inconceivable for a variety of reasons~\cite{McDonald-SID};
some being either technical, others related to secrecy concerns, such as compartmentalization by need-to-know principles.
However, we would like to point out that any concerns related to security can be addressed by transforming these data into forms that are not exploitable
by potential adversaries. This tranformation would most likely include omitting data fields, and dowgrading high pecision military data to civil needs.
A historic example for such an access-by-transformation is the  Global Positioning System (GPS):
During the 1990s, GPS employed a degradation feature called Selective Availability that intentionally degraded civilian accuracy on a global basis~\cite{GPSAccuracy}.

We therefore suggest, as others before~\cite{Dean-Willis-podcast-2020,Mellon2022Feb},
that resources from the US Space Surveillance Network categorized as ``uncorrelated targets'' in the broadest sense---that is, not categorized as correlated target,
or identifiable military threat such as ICBMs or enemy airplanes---are made openly accessible.
These ``anomalies'' might be delivered in a possibly downgraded form for civilian use;
in particular for the analysis of tracks that might yield information about future feasible technologies.

\subsubsection{Types of sensors: optical, radar, infrared and acoustic}

What sort of data can be expected; in particular, from what sources?
How are these sources integrated and channelled into what organizational (sub)units?

Let us concentrate on the United States, and first consider ground based sensors ``looking upwards and skywards''.
In the United States (US) the Strategic Command~\cite{jp3_14Ch1} has assigned the responsibility for Space Situational Awareness (SSA)
to its Joint Space Operations Center (JSpOC)  for the Joint Functional Component Command - Space (JFCC Space), all
at Vandenberg Air Force Base~\cite{Wasson2011amos.conf}.
The data is also supplied to the
National Air and Space Intelligence Center (NASIC) situated on Wright-Patterson Air Force Base outside of Dayton, Ohio~\cite{Bruck2014Jun}.
JFCC Space is responsible for tracking all man-made objects in orbit. The center receives on-orbit positional data, known as element sets.
All sensors provided by NASA and the Air Force are combined into the United States Space Surveillance Network (SSN).
Data are
supports United States Strategic Commands USSTRATCOM mission Space Surveillance awareness SSA.
Let us first mention United States ground based sensors, followed by space based satellite ones.

\begin{itemize}
\item[(I)] The following electro-optical sensors are electronic detectors converting light, or a change in light, into electronic signals:
\begin{itemize}
\item[(i)]
the Ground-Based Electro-Optical Deep Space Surveillance System (GEODSS)~\cite{GEODSS},
The three operational GEODSS sites are: Detachment 1, Socorro, New Mexico, USA; Detachment 2, Diego Garcia, British Indian Ocean Territory;
and Detachment 3, Maui, Hawaii. According to the Air Force~\cite{GEODSS}, ``the GEODSS system can track objects as small as a basketball more than
20,000 miles away and is a vital part of the Air Force Space Command's (AFSPC's) space surveillance network.''
[Previously the Mor\'on Optical Space Surveillance System (MOSS) has been an electro-optical surveillance system in Spain
intended to be a gap filler operating in concert with GEODSS.]
\item[(ii)]
the Space Surveillance Telescope (SST) in Exmouth, Western Australia.
\end{itemize}

\item[(II)] The following ground based (mostly phased array) radar systems comprise:
\begin{itemize}
\item[(i)]
GLOBUS II and III radar  in Norway,
\item[(ii)]
AN/FPS-85 Space Track Radar, a phased array radar in Florida,
\item[(iii)]
AN/FPS-133 Air Force Space Surveillance System, also known as the Space Fence,
is located at Kwajalein Atoll in the Marshall Islands,
\item[(iv)]
auxiliary phased array radar sensor systems,
such as the
Solid State Phased Array Radar System (SSPARS) comprising of  AN/FPS-132 Upgraded Early Warning Radar (UEWR) systems
deployed at multiple sites,
the AN/FPS-108 Cobra Dane in Alaska,
and the AN/FPQ-16 Perimeter Acquisition Radar Characterization System (PARCS) in North Dacota~\cite{ContributorstoWikimediaprojects2021Dec}.
\end{itemize}
\item[(III)]
Both ground based
electro-optical as well as radar sensors are augmented by satellites
[previously the Space Based Visible (SBV) sensor of the  Midcourse Space Experiment (MSX)]
of the Space Based Space Surveillance (SBSS)
system of ``pathfinder'' satellites.
\end{itemize}

Let us again concentrate on the United States, and second, space based satellite systems of sensors.
The Space Based Infrared System (SBIRS) and its predecessor, the Defense Support Program (DSP) satellites~\cite{Richelson2012},
is a network of satellites operating in medium-earth, highly elliptical,
and geosynchronous orbits that together provide continuous global coverage of infrared energy sources~\cite{Mellon2022Feb}.

Allegedly the KH-11 KENNEN from Block 2 on equipped with infrared sensors,
is a type of electro-optical reconnaissance satellite first launched by the American National Reconnaissance Office in December 1976,
that lead to the detection of ``orange orbs'' in the frequency range of about 5 THz, wavelength 600 nm,
flying in formation~\cite[t=1620s=27:00m,t=2270s=37:50m]{Ramirez-Greenwald-22}.

Data of the space based sensors are reported to
Air Force Space Command's  21st Operations Group, 21st Space Wing at Wright-Patterson Air Force Base~\cite[t=2340s=39:00m]{Dean-Willis-podcast-2020}
that became part of the United States Space Force's (USSF) Space Delta 4 (DEL 4) at Buckley Space Force Base, Buckley Garrison, Colorado~\cite{BSFB}.
At least some of the data might, for analysis, also go to NASIC at Wright-Patterson Air Force Base~\cite{NASIC,DeanPaul-NASIC}.
Teams of space and counterspace analysts at NASIC
will be transferred to The Space Force Intelligence Activity (SFIA), which in turn is an
``interim operational construct to facilitate the process of the National Space Intelligence Center establishment''~\cite{SFIA}.

The Global Infrasound Acoustic Monitoring Network
comprised of 60 stations in 35 countries that monitor low-frequency pressure waves in the atmosphere~\cite{Mellon2022Feb}.

There may be some interesting signal characteristics, as McDonald mentioned~\cite{McDonald-SID} a 2800~MHz $\approx$ 3~GHz signal,
a common frequency for S-band search termed ``magic'' by Ramirez~\cite[t=2943s=48:03m]{Ramirez-Willis-Dec21},
encountered by an electronic countermeasure (ECM) gear equipped USAF Boeing RB-47 Stratojet in September~1957.
This, as noted by Ramirez~\cite[t=3065s=51:05m]{Ramirez-Willis-Dec21}, is in the frequency range of 3-4~GHz, wavelength 7.5-10~cm,
of the [USS PRINCETON (CG 59)] AEGIS AN/SPY-1 radar~\cite{MoenVM208132522013.36586}.

Regarding the collection and access of all sorts of ``uncorrelated tracks'' in the United States
it might not be unreasonable to speculate that, on the one hand, the USAF and USSF and their various branches
(e.g., The National Intelligence Manager for the Air Domain (NIM-AD)~\cite{NIM-AD} or NASIC/SFIA)
have, beginning with the end of the second world war and possibly earlier, continuously and without interruption,
accumulated a huge body of data classified as (top) secret and compartimented by  need-to-know principles.
On the other hand, there have been
``ad hoc'' congressional initiatives for (more) oversight that have either initiated or stimulated AWWSAP and AATIP~\cite{Lacatski-2021},
as well as enquiries by the  Defense Intelligence Agency  (DIA)
and like minded attempts by the Department of Homeland Security (DHS)~\cite[Chapters~11,16]{Lacatski-2021};
all of which might not have been able to tap into these aforementioned USAF and USSF data~\cite{Mellon2022Feb}.
The success of congressional initiatives like the Unidentified Aerial Phenomena Amendment included in the final fiscal year 2022 National Defense Authorization Act
(FY22 NDAA)~\cite{Gillibrand-FY22NDAA} and the newly created Airborne Object Identification and Management Synch (AOIMSG)~\cite{AOIMSG}
thus depend on whether USAF and USSF will participate.

We suggest, as others before~\cite{Dean-Willis-podcast-2020,Mellon2022Feb}, to go one further step, and make pertinent data publicly available.
Resources from the US Space Surveillance Network should be provided as long as they are categorized as
``uncorrelated targets'' in the widest sense if they do not presenting any conventional threats to homeland security.
Conventional threats will be most interesting for USAF and USSF purposes but are of no interest in this context.
These ``anomalies'' might be delivered in a possibly downgraded form for civilian use;
in particular for the analysis of tracks that might yield information about future feasible technologies.

\section{On the possible lack of autonomy and self-determination}

Even another speculative hypothetical worriment must be seen in a wider context:
there might be potent, even dominant, ``other (outer) parties'' that control and constrain detection and disclosure,
as this might be unfavorable to their intents and interests.
However, conformity to such objectives might get increasingly onerous as technology progresses.
In one way or another, space above (and below) ground level will open up to general inspection.
This might be accompanied by sobering acknowledgements
not dissimilar to the Copernican revolution~\cite{DeLongeHartley-SM1,DeLongeLevenda-Gods,DeLongeHartley-SM2,DeLongeLevenda-Men,Dolan-alien-agendas,Lacatski-2021}.

We end with a related issue, as expressed in the Japanese-Polish movie ``Avalon''~\cite{Avalon2001}:
``What is the better game, one of which you think you can leave but can't,
or one that looks impossible to leave but an exit always exists?''
Is it always better to know, even if that knowledge is harmful and causes pain?
Probably not.
And yet, as noted by Goethe, ``No one is more a slave than he who thinks he is free without being so.''

\ifx\revtex\undefined

\funding{
}

\acknowledgments{This paper is dedicated to the memory of Dr. Kurt Toman (1921-2005), a graduate of the TU Wien (1949) and the University of Illinois at Urbana (1952),
and Chief of the Ionospheric Properties Branch, Upper Atmosphere Physics Laboratory, Air Force Cambridge Research Laboratories (AFCRL), USA.


I kindly acknowledge discussions with and suggestions by Andreas Neubacher.}

\conflictsofinterest{The author declares no conflict of interest.
The funders had no role in the design of the study; in the collection, analyses, or interpretation of data; in the writing of the manuscript, or in the decision to publish the~results.}

\else

\begin{acknowledgments}
This paper is dedicated to the memory of Dr. Kurt Toman (1921-2005), a graduate of the TU Wien (1949) and the University of Illinois at Urbana (1952),
and Chief of the Ionospheric Properties Branch, Upper Atmosphere Physics Laboratory, Air Force Cambridge Research Laboratories (AFCRL), USA.


I kindly acknowledge discussions with and suggestions by Paul Dean, Richard G. Hopf, Andreas Neubacher, and participation in a new form of ``Invisible College''~\cite{Vallee-IC}.


The author declares no conflict of interest.
\end{acknowledgments}

\fi

\ifx\revtex\undefined

\end{paracol}
\reftitle{References}


 \externalbibliography{yes}
 \bibliography{svozil,ufo}

\else


%

\fi
\end{document}